\documentclass[12pt]{article}
\usepackage{epsfig}
\usepackage{rotating}
\title{Non-universality of commonly used correlation-energy 
density functionals}

\author{ Jacob Katriel\thanks{\noindent 
Permanent address: Department of Chemistry, Technion, Haifa 32000, Israel. 
\hfill\break  email: jkatriel@tx.technion.ac.il},$\;$
Sudip Roy\thanks{\noindent email: suro001@rz.uni-saarland.de}$\;\;$  and 
Michael Springborg\thanks{\noindent email: m.springborg@mx.uni-saarland.de}\\   
{\small{ Physikalische und Theoretische Chemie, Universit\"at des Saarlandes}}\\ 
{\small{66123 Saarbr\"ucken, Germany }}  }

\date{ \today}  

\begin{document}
\setcounter{page}{1}

\maketitle

\begin{abstract}
The correlation energies of the helium isoelectronic sequence and 
of Hooke's atom isoelectronic sequence have been evaluated using an 
assortment of local, gradient and meta-gradient density 
functionals. The results are compared with the exact correlation energies,
showing that while several of the more recent density functionals reproduce
the exact correlation energies of the helium isoelectronic sequence rather 
closely, none is satisfactory for Hooke's atom isoelectronic sequence. It 
is argued that the uniformly acceptable results for the helium sequence can 
be explained 
through simple scaling arguments that do not hold for Hooke's atom sequence,
so that the latter system provides a more sensitive testing ground for 
approximate
density functionals.
This state of affairs calls for further effort towards formulating 
correlation-energy density functionals that would be truly universal 
at least for spherically-symmetric two-fermion systems.
\end{abstract}

\noindent
keywords: Correlation energy, density functionals, helium, 
Hooke's atom. 

\vfill\eject

\section{Introduction}

Density functional theory (dft) can be implemented in a variety of ways,
all of which have their roots in the pre-Hohenberg-Kohn era.
A fully consistent implementation requires the formulation of 
density functionals for the kinetic energy as well as for the exchange and 
correlation energies, none of which can be expressed in terms of the 
one-particle density in a straightforward way. The Kohn-Sham formulation 
depends on the construction of a local one-body potential that incorporates
the effect of interelectronic repulsion, giving rise 
to a ground state density that coincides with that derived from the 
solution of the exact many-body hamiltonian.
Finally, an approximate approach, 
based on the fact that the Hartree-Fock density is fairly close to the
exact one, proposes that the exact ground state energy be evaluated 
by adding to the Hartree-Fock energy a correlation energy, calculated in terms
of the Hartree-Fock density by means 
of an appropriate density functional.

Since the ultimate correlation-energy density functional, whose
existence is guaranteed by the Hohenberg-Kohn theorem, remains elusive,
the development and testing of ever improved density functionals keeps being 
vigorously pursued. 
For an extensive recent study see Kurth, Perdew and Blaha \cite{Kurth}.
In this vein, Chakravorty {\it et al.}\ \cite{Chak} and 
Jarzecki and Davidson \cite{Jarz} compared the exact and dft correlation 
energies across the helium (and other) atomic isoelectronic sequences. 
In the former article the exact correlation energy for the ground state of 
the helium isoelectronic sequence (as well as for all other atomic 
isoelectronic sequences with up to 18 electrons) is compared with the
values obtained using six density functionals (i.e., LYP \cite{LYP}, 
CS \cite{Colle}, VWN-SPP \cite{SPP}, PZ \cite{PZ}, B88 \cite{Be}, 
LC \cite{LC} --- the acronyms will be explained below).
Of these, only the LYP correlation energies are uniformly close to the 
exact values over the range of nuclear charges examined.
B88 grows to become twice as large as the exact correlation
energy, and all the others seem to diverge for large Z. The latter article
extends this study, considering P86 \cite{P86}, LYP \cite{LYP} (again), and 
PW91 \cite{Perdew92} as well as various
combinations of these with exchange functionals. While, as noted above, 
the correlation energy obtained with the LYP functional (evaluated for the 
Hartree-Fock density) follows the exact correlation energy rather closely, 
this is not the case if the same functional is combined with
Becke's exchange functional. On the other hand, the combination BP86
(Becke's exchange with the P86 correlation functional) exhibits a rather
close agreement with the exact correlation energies for the helium 
isoelectronic sequence.
The overall conclusion of these authors is ``all DFT functionals, [...]
reveal problems in descriptions of the energy as a function of nuclear
charge'' \cite{Jarz}.
Correlation energies were also evaluated by Umrigar and Gonze \cite{UG} for
several members of the helium isoelectronic sequence, in terms of both
local and GGA-type density functionals.  

The high $Z$ limit of the helium (and other) isoelectronic sequences
within dft was carefully examined by Whittingham and Burke \cite{WB},
who pointed out that some correlation-energy functionals do not scale correctly
to the high-density limit. In particular, the local-density approximation
diverges logarithmically at large $Z$, and the Perdew-Wang (PW91) functional
\cite{Perdew92} fails as well. They note that PBE \cite{PBE} was designed
to correct for this failure.

The over-estimate of atomic correlation energies by local density
approximations based on the homogeneous electron gas has been analyzed 
by Tong \cite{Tong}, who attributed it to the continuous excitation 
spectrum of the latter, infinitely extending, system, that allows for a ``soft''
response to 
the internuclear repulsion. A modification of the local density approximation, 
in which an energy gap is introduced, was recently studied by 
Rey and Savin \cite{RS}. Their results suggest that the inhomogeneity needs
to be introduced more explicitly.

An analysis of the properties that correlation-energy density functionals
should satisfy in order to present a valid behavior along isoelectronic
sequences was presented by Staroverov {\it{et al.}} \cite{SSPTD}, who, as 
Whittingham and Burke \cite{WB}, studied the high $Z$ limit of some two-electron
atomic ions. Furthermore, an
interesting methodological discussion of the construction of density
functionals was recently presented by Perdew {\it{et al.}} \cite{PRTSSC}.
These authors make a strong case for nonempirical functionals and specifically
recommend the PBE \cite{PBE} and the TPSS \cite{TPSS} functionals.

Hooke's atom consists of two ``electrons'' that repel one another like
ordinary electrons but are bound to the ``nucleus'' via harmonic potentials.
Its separability, in center-of-mass and relative coordinates, was pointed out 
by Kestner and Sinanoglu \cite{Kestner}, and the fact that for certain
values of the force constant it is analytically solvable
was noted by Kais {\it{et al.}} \cite{KaisandLevine} and, more systematically,
by Taut \cite{Taut}.
Several studies for Hooke's atom have been performed, aimed at deriving
exact density functionals or Kohn-Sham type exchange-correlation
potentials [23-25],
most notably by Burke {\it{et al.}} [26-30]
and Ludena {\it{et al.}} \cite{Ludena}.
However, a systematic study over a broad range of force constants has not been
presented, to the best of our knowledge, with the exception of the
local-density calculations that were carried out by Huang and Umrigar \cite{HU}.

In the present paper we examine an assortment of correlation-energy
density functionals proposed over the last seventy(!) years. 
Virtually each one of them had been tested for the helium atom, and
most of them had been tested for a range of atomic ions isoelectronic 
with helium. Needless to say, many of them have been applied, with varying
degrees of success, to the study of heavier atoms as well as molecules, 
surfaces and solids, and share the credit for the immense impact of dft in 
all these areas.

The Hohenberg-Kohn theorem implies that the
ground-state correlation energy is a universal functional of the ground 
state density. The examination of the extent to which the commonly used
correlation-energy density functionals satisfy this property is the principal
aim of the present paper. This is carried out by considering the performance 
of the presently selected set of correlation energy density functionals for 
Hooke's atom isoelectronic sequence.  
Since these density functionals were originally 
developed for Coulombic (atomic and molecular) systems, their performance
for Hooke's atom provides a severe test of their universal validity.
Moreover, Hooke's atom remains bound over the whole range, 
$0<k<\infty$, of the force constant. This fact allows
correlation to be examined not only in the high $k$ (weak correlation) limit
but also in the low $k$ (strong correlation) limit. The latter limit is not
available for atomic systems, that become unbound below a certain critical
nuclear charge. This property of Hooke's atom isoelectronic sequence 
is elaborated upon below.

Another model system with similar properties is that of a two interacting 
particles confined to a spherical volume. Here, the radius of the sphere provides
a parameter that, upon variation, can be used to change the system from the low correlation
to the strong correlation limit. This system was recently studied in detail by Jung {\it et al.}
\cite{JGAG} who also compared the exact exchange and correlation energies with those obtained
by some few approximate functionals. Thus, the combined results of their and our study provide
detailed insight into the accuracies of currently applied approximate density functionals
and, consequently, may be used in improving those.

Ultimately, the physical reason behind the existence of correlation
effects is the electron-electron interactions. Thus, the total correlation
energy can be written in terms of a double-integral over two position 
coordinates, involving the exact and the Hartree-Fock first and second 
order reduced density matrices,
\begin{equation}
E_c=\int\int \rho(\vec r_1)\rho(\vec r_2) 
         \tilde\epsilon_c(\vec r_1,\vec r_2)d^3\vec r_1d^3\vec r_2,
\label{mcs01}
\end{equation}
where $\tilde\epsilon_c(\vec r_1,\vec r_2)$ is a unique correlation-energy 
density
in a six-dimensional $(\vec r_1,\vec r_2)$ space. On the other hand,
in density functional theory it is common practice to write the correlation
energy as
\begin{equation}
E_c=\int \rho(\vec r)\epsilon_c(\vec r)d^3\vec r.
\label{mcs02}
\end{equation}
The transformation from Eq.\ (\ref{mcs01}) to Eq.\ (\ref{mcs02}) is not
unique. Therefore, different approximate expressions for $\epsilon_c(\vec r)$ may
look very different but nevertheless give very close total correlation energies $E_c$ \cite{TP}.
Thus, since the approximate correlation-energy functionals most often are given in
terms of approximate forms for $\epsilon_c(\vec r)$, it is not possible to compare
those directly and thereby obtain information on their performance. 

Instead, only the total correlation energy is of physical relevance. Therefore, by
considering the total correlation energy for continuous classes of systems detailed
information on the performance of approximate density functionals can be 
obtained,
that ultimately can be used in assessing their performance as well as in improving them.
The helium isoelectronic sequence provides such a class of systems, as 
also is the case
for Hooke's atom sequence. However, as briefly pointed out above and as we 
shall demonstrate more carefully below, the former sequence is less
general than what initially may be assumed, making it less informative 
for studies of 
approximate energy functionals. On the other hand, as we shall also demonstrate, 
Hooke's atoms provide a class of systems for which the members are 
``less similar''
than is the case for the helium sequence, making a detailed study of this 
sequence highly relevant. 

We finally add that correlation energy is defined differently in wave function methods
and in density functional theory. Since the numerical differences are not large and
not will affect any of the conclusions of the present work, we shall not discuss this
distinction further.

\section{The spherical two-electron systems}

The ground state of a spherically symmetric two particle system with the Hamiltonian 
\begin{equation}
\label{eq:hamiltonian}
{\cal H}=-\frac{1}{2}\left(\nabla_1^2+\nabla_2^2\right)
             +v(r_1)+v(r_2)+\frac{1}{r_{12}} \, ,
\end{equation}
in which the (local) single-particle potential $v(r)$ is sufficiently 
attractive, 
is a bound singlet; hence, it satisfies $\rho_{\alpha}=\rho_{\beta}$, where 
$\rho_{\alpha}$ and $\rho_{\beta}$ are the spin up and spin down densities, 
respectively; i.e., the spin polarization parameter 
$\zeta=\frac{\rho_{\alpha}-\rho_{\beta}}{\rho_{\alpha}+\rho_{\beta}}$ vanishes.
The spatial part of the corresponding Hartree-Fock wavefunction is of the form
$\Phi_{HF}(\vec{r_1}, \vec{r_2})=\phi(r_1)\phi(r_2)$, and
the electron density is given in the form
$\rho(r)=2\phi(r)^2 \, .$
It will be useful to note that $|\nabla\rho(r)|=\frac{d\rho}{dr}$, 
$\left(\frac{d\phi}{dr}\right)^2
         =\frac{1}{8\rho}\left(\frac{d\rho}{dr}\right)^2$,  
and $\nabla^2\rho=\frac{d^2\rho}{dr^2}+\frac{2}{r}\frac{d\rho}{dr}$.
Moreover, in the present context 
$\left(\frac{d\phi}{dr}\right)^2
            =\frac{1}{8\rho}\left(\frac{d\rho}{dr}\right)^2$.
For each of the density functionals considered below we provide the (sometimes
considerably simplified) expression 
appropriate for the family of spherically symmetric two-particle systems 
presently studied. One reason for doing so is that the expressions 
for the various density functionals are scattered in the literature, in many
instances the expression for a particular functional requiring consultation of
several different sources whose notation is occasionally not fully consistent.
The record is held by TPSS, the specification of which requires consultation of
ref. \cite{TPSS}, \cite{PKZB}, \cite{PBE}, and \cite{Perdew92}.

In our case we have 
\begin{equation}
v(r)=\cases{\frac{-Z}{r} & for the helium sequence \cr
            \frac{1}{2}kr^2 & for Hooke's atom. \cr}
\end{equation}

The correlation energy is given in terms of an expression of the form
$$E_c=\int_0^{\infty}4\pi r^2\rho(r)\epsilon_c[\rho]dr$$ 
where $\epsilon_c[\rho]$ is a correlation energy density whose value at 
$r$ may depend on the values of $\rho$, $\frac{d\rho}{dr}$ as well as 
$\frac{d^2\rho}{dr^2}$ at $r$.
The dependence of the various functionals on the density is usually expressed
via the Wigner-Seitz radius (the radius of a sphere containing one particle 
with uniform density $\rho$) 
$$r_s=\left(\frac{3}{4\pi\rho}\right)^{1/3} \; .$$

We follow the classification of $\epsilon_c[\rho]$ into
\begin{itemize}

\item
Zero-order approximation (0OA), depending on $\rho$ only,

\item
First-order approximation (1OA), depending on $\frac{d\rho}{dr}$ 
as well, 

\item Second-order approximation (2OA), depending also on $\frac{d^2\rho}{dr^2}$.  

\end{itemize}

We note that this classification is closely related to the common classification 
into local density approximations (LDAs), 
generalized-gradient approximations (GGAs) and meta-GGAs. However, by
restricting the classification according to the behavior for a spherical 
two-electron system, for a given functional we sometimes arrive at an expression that belongs 
to a lower class, in the classification specified above, than the general case. 
Our classification here applies to this reduced expression.

Both as an additional check of the computations and to gain further insight 
into the behavior of the various density functionals considered, we
examine the weak correlation limit ($Z\rightarrow\infty$
or $k\rightarrow\infty$, for the Helium and Hooke's atom isoelectronic
sequences, respectively) of the corresponding correlation energies.
This can be done by using the asymptotic densities
$$\rho({\rm He})=\frac{2 Z^3}{\pi} \exp(-2 Z r),\;\;\; Z\rightarrow\infty$$ and
$$\rho({\rm Hooke})=2\left(\frac{\omega}{\pi}\right)^{3/2}\exp(-\omega r^2), \;\;
\omega=\sqrt{k}, \;\;\; k\rightarrow\infty \, .$$
In fact, the density, correct to first order in the interelectronic
repulsion, can be written for the Helium isoelectronic sequence by replacing
the nuclear charge $Z$ by the screened nuclear charge $Z_{\rm eff}=Z-5/16$.
Similarly, for Hooke's atom the variational principle yields for the trial
function
$\left(\frac{\omega_{\rm eff}}{\pi}\right)^{3/4}\exp(-\frac{\omega_{\rm eff}}{2} r^2)$
the equation
$\omega_{\rm eff}^2+\frac{1}{3}\sqrt{\frac{2}{\pi}}\omega_{\rm eff}^{3/2}-\omega^2=0$,
where $\omega=\sqrt{k}$. The solution, to first order, can be written in
the form
$\omega_{\rm eff} \approx
        \left(\sqrt{\omega}-\frac{1}{12}\sqrt{\frac{2}{\pi}}\right)^2$.
In fact, to study the weak correlation limit it is convenient to use the
scaled radial coordinate $\tilde{r}=Z_{\rm eff} r$ or
$\tilde{r}=\omega_{\rm eff}^{1/2} r$, for the Helium and Hooke's atom
isoelectronic sequences, respectively. In terms of these coordinates
$\rho(r) 4\pi r^2 dr$ becomes $8\exp(-2\tilde{r}) \tilde{r}^2 d\tilde{r}$
or $\frac{8}{\sqrt{\pi}}\exp(-\tilde{r}^2) \tilde{r}^2 d\tilde{r}$,
respectively. In terms of these scaled coordinates the expressions
$\frac{d\rho}{dr} \rho^{-4/3}$, $\frac{1}{r}\frac{d\rho}{dr} \rho^{-5/3}$ and
$\frac{d^2\rho}{dr^2} \rho^{-5/3}$, that appear in several of the GGA and
meta-GGA density functionals, obtain $Z$- (or $k$-) independent forms in
the weak correlation limit. The weak correlation limit for the various density 
functionals examined below should be compared with the exact values 
\cite{Brown}
$$E_c({\rm He}) \approx -0.046663 + \frac{0.009739}{Z}$$
        and
$$E_c({\rm Hooke}) \approx -0.049703 + \frac{0.009369}{k^{1/4}} \, .$$

\section{Zero order approximations}

\subsection{Wigner's functional}

Wigner's first exploration of the correlation energy density as a means for
estimating correlation energies in many-electron systems dates back to
1934 \cite{Wigner34}.
Four years later he proposed the functional \cite{Wigner38}
\begin{equation}
\label{eq:Wigner}
\epsilon_c^W(r_s) = -\frac{0.44}{7.8+r_s} \; .
\end{equation}
In the weak-correlation limit (i.e., when either $Z\rightarrow \infty$ or
$k\rightarrow\infty$) Wigner's correlation energy for a two-electron system
becomes $-\frac{2\cdot 0.44}{7.8}=-0.113$, which is roughly twice as large
as the corresponding (rather similar) limiting values of the helium and
Hooke's atom isoelectronic sequences. A closer examination yields
$E_c^W({\rm He}) \approx -0.113 + \frac{0.035}{Z}$  and
$E_c^W({\rm Hooke}) \approx -0.113 + \frac{0.023}{k^{1/4}}$.
These expressions are easily obtained analytically, noting that in this limit
$\epsilon_c^W \approx -\frac{0.44}{7.8} + \frac{0.44}{7.8^2}r_s$.
In the strong correlation limit $\epsilon_c^W\approx -\frac{0.44}{r_s}$.
            
\subsection{The Gunnarsson-Lundqvist functional \cite{GL}}

\begin{equation}
\label{eq:GL}  
\epsilon_c^{GL} = -0.0333\left[ (1+x_p^3)\log(1+\frac{1}{x_p})
                        +\frac{x_p}{2}-x_p^2-\frac{1}{3} \right]
\end{equation}  
where $x_p=\frac{r_s}{11.4}$.
Since for $r_s\rightarrow 0$ one has $\epsilon_c^{GL}\approx 0.0333\log(r_s)$,
it follows, using the asymptotic densities specified above, that
$$E_c^{GL}({\rm He},Z\rightarrow\infty)\approx -2\cdot 0.0333\log(Z)-0.09506 \, ,$$
and
$$E_c^{GL}({\rm Hooke},k\rightarrow\infty)\approx 
            -\frac{0.0333}{2} \log(k) -0.116  \, .$$
The leading asymptotic terms for the helium and Hooke's sequences are related
to one another by the substitution of $k^{\frac{1}{4}}$ for $Z$,
which is consistent with McWeeny's correspondence \cite{McW}.
In the low density (strong correlation) limit, $r_s\rightarrow\infty$, hence 
$\epsilon_c^{GL}\approx -\frac{0.285}{r_s}$.
                
\subsection{The Brual-Rothstein functional \cite{BR}}

This correlation energy density
\begin{equation}
\label{eq:BR}
\epsilon_c^{BR}=-\frac{1}{9.81+21.437\rho^{-1/3}}
                    = -\frac{0.0289}{0.284 + r_s}
\end{equation}
has the same form as Wigner's functional, eq. \ref{eq:Wigner}.
The parameterization has been chosen to yield an exact correlation energy 
for the helium atom.

Just like Wigner's functional, the weak correlation limit yields
a common correlation energy for the hydrogenic and Hooke's sequences,
$E_c^{BR}=-0.204$.

\subsection{The Perdew-Wang parameterization \cite{Perdew92}}

This is the up-to-date local density approximation, that we write in the form 
\begin{equation}
\label{eq:LDA}
\epsilon_c^{unif}(r_s)=-2c_0(1+\alpha_1 r_s)
\log\left[ 1+ \frac{1}{{\tilde{\beta}}_1r_s^{1/2}+{\tilde{\beta}_2}r_s
   +{\tilde{\beta}_3}r_s^{3/2}+{\tilde{\beta}_4}r_s^2}\right] \, ,
\end{equation}
where
\begin{eqnarray*}
c_0 &=& 0.031091 \\
\alpha_1 &=& 0.21370 \\
{\tilde{\beta}}_1 &=& 2c_0\beta_1 =0.47232 \\
{\tilde{\beta}}_2 &=& 2c_0\beta_2 =0.22308 \\
{\tilde{\beta}}_3 &=& 2c_0\beta_3 =0.10187 \\
{\tilde{\beta}}_4 &=& 2c_0\beta_4 =0.030652 \, . 
\end{eqnarray*}
$\beta_i$, $i=1,2,3,4$, (along with $c_0$ and $\alpha_1$) are given 
in \cite{Perdew92}. In the weak correlation limit ($r_s\rightarrow 0$) one
obtains 
$\epsilon_c^{unif}(r_s)\approx 2c_0\log({\tilde{\beta}}_1 r_s^{1/2})$, hence
$$E_c^{PW}({\rm He})\approx -2 c_0\log(Z)-0.05143 \, ,$$
and
$$E_c^{PW}({\rm Hooke})\approx -\frac{c_0}{2} \log(k) -0.07066 \, .$$
These expressions agree with the computed correlation energies, presented 
in tables 1 and 4, respectively. In the strong correlation limit, 
$r_s\rightarrow\infty$, we obtain
$\epsilon_c^{unif} \approx -\frac{0.434}{r_s}$, which is rather close to the 
corresponding limit of Wigner's correlation-energy density functional, 
eq. \ref{eq:Wigner}.

\section{First-order approximations}

\subsection{The Perdew86 functional \cite{P86}}

Define
$$A = 0.002568 + 0.023266 r_s + 7.389\cdot 10^{-6} r_s^2$$
and
$$B= 1+ 8.723 r_s + 0.472 r_s^2 + 0.07389 r_s^3 \; .$$
Then 
$C=0.001667 + \frac{A}{B}$ and
$$\Phi = \frac{-0.0008129}{C} \cdot \frac{d\rho}{dr} \rho^{-7/6} \; .$$

The correlation energy density for the homogeneous electron gas is parameterized
as follows \cite{PZ}:
For $r_s<1$:
$$\epsilon_c^{(0)} = 
      0.0311 \log(r_s) -0.048 + 0.0020 r_s \log(r_s) -0.0116 r_s \, ,$$
[$E_c^{(0)}({\rm He},Z\rightarrow\infty) =  -2\cdot 0.0311\log(Z)$;
$E_c^{(0)}({\rm Hooke},k\rightarrow\infty) = -\frac{0.0311}{2}\log(k)$]

\noindent
and for $r_s>1$:
$$\epsilon_c^{(0)}= \frac{-0.1423}{1 + 1.0529 r_s^{1/2} + 0.3334 r_s } \, .$$
[For $r_s\rightarrow\infty$, $\epsilon_c^{(0)}\approx -\frac{0.427}{r_s}$].

In terms of all the above, the correlation energy density is
$$\epsilon_c^{P86} = \epsilon_c^{(0)} 
      +C \exp(-\Phi) \left(\frac{d\rho}{dr}\right)^2 \rho^{-7/3} \; .$$
In the $Z\rightarrow\infty$ (or $k\rightarrow\infty$) limit the non local term
vanishes, because in this limit $\Phi\sim Z^{1/2}$ 
($\Phi\sim k^{1/8}$, respectively) so $\exp(-\Phi)\rightarrow 0$.

\subsection{Becke's functional \cite{Be}}

Define
$B= \left( \frac{d\rho}{dr}\right)^2  \rho^{-8/3} $.
Then
$$ E_x=-\rho^{1/3} \left( \frac{3}{4} \left(\frac{3}{\pi}\right)^{1/3}
         + 0.00375\cdot 2^{1/3} B \Big(1 + 0.007\cdot 2^{2/3} B \Big)^{-4/5}
                               \right) \; .$$
Now define $z=-\frac{0.63}{E_x}$.
The expression for the correlation energy density is
$$\epsilon_c^{B88} = -0.2 \rho z^2 \left(1- \frac{\log(1+z)}{z} \right) \, . $$
In the weak correlation limit it is found that $z\rightarrow 0$, hence
$\epsilon_c^B \approx -0.1\rho z^3$. Numerical integration over the asymptotic
densities yields $E_c^{B88}({\rm He},Z\rightarrow\infty)=-0.07445$ and
$E_c^{B88}({\rm Hooke},k\rightarrow\infty) = -0.22176$.
            
\subsection{The Wilson-Levy functional \cite{WL}}

This functional,
\begin{equation}
\label{eq:WL}  
\epsilon_c^{WL}=\frac{-0.74860
    + 0.06001 \left|\frac{d\rho}{dr}\right| \rho^{-4/3} }
     {3.60073+2.2679\left|\frac{d\rho}{dr}\right|\rho^{-4/3} +r_s} \, ,
\end{equation}  
is a non-local generalization of Wigner's functional.
Numerical integration over the
asymptotic densities yields $E_c^{WL}({\rm He},Z\rightarrow\infty)=-0.04803$
and  $E_c^{WL}({\rm Hooke},k\rightarrow\infty)=-0.07817$.
                
\subsection{The Perdew, Burke and Ernzerhof functional \cite{PBE}}

For a spherically symmetric singlet state this correlation-energy density 
is of the form
\begin{equation}
\label{eq:PBE}
\epsilon_c^{PBE} = \epsilon_c^{unif}(r_s) + H(r_s,t) \; .
\end{equation}
$\epsilon_c^{unif}(r_s)$ is given by eq. \ref{eq:LDA}, and
$$H(r_s,t)=\gamma\log\left\{1+\delta
 t^2\left[\frac{1+At^2}{1+At^2+(At^2)^2}\right]\right\} \; ,$$
where $t=\frac{1}{4}\left(\frac{\pi}{3}\right)^{1/6} \rho^{-7/6}
          \frac{d\rho}{dr}$ and
$A=\delta\left[\exp
         \left(-\frac{\epsilon_c^{unif}}{\gamma}\right)-1\right]^{-1}$.
Here, $\gamma=c_0$ ({\it{cf.}} eq. \ref{eq:LDA}) and $\delta=2.1461$. 
The $Z\rightarrow\infty$ logarithmic leading term of 
$\epsilon_c^{unif}(r_s)$ is canceled by an equal term with opposite sign in
$H(r_s,t)$.
A recent comparative study \cite{Kurth} suggests that this correlation energy
density functional is superior to all other density functionals
considered in that study.
The weak correlation limit requires numerical integration over the asymptotic 
densities, yielding $E_c^{PBE}({\rm He},Z\rightarrow\infty)=-0.04789$, 
in agreement 
with \cite{PBE} and \cite{SSPTD}, and $E_c^{PBE}({\rm Hooke}, k\rightarrow\infty)=-0.08131$.

\subsection{The Filatov-Thiel density functional}

Filatov and Thiel proposed two closely related functionals, that we
shall designate FT97 (cf. \cite{FT97a,FT97b}) and FT98 (cf. \cite{FT98}).
Whereas FT97 belongs in the present subsection, FT98 contains a dependence
on the Laplacian of the density and belongs in the following subsection.

FT97 ``[...] is based on a model of the Coulomb hole distribution function
and accurately describes the correlation energy of the uniform electron gas
and the atoms He to Ar'' \cite{FT97b}.
Define
$$y_0 = \frac{1}{(18\pi)^{2/3} } \rho^{-8/3} \left(\frac{d\rho}{dr}\right)^2$$
and let
\begin{equation}
\label{eq:FF97}
F_{97}=\frac{1+1.622199 y_0 + (0.489958 y_0)^2}
                         {\sqrt{1+1.094530  y_0/r_s} }
                          \exp[-(0.489958 y_0)^2] \; .
\end{equation}
Let
$$\mu= \frac{0.02940 r_s}
{\left\{\left[1 + 0.370365 \exp(-0.10018 r_s^{4/5})\right]
                F_{97}\right\}^2} \; .$$
and
$$f(\mu)=-\exp(\mu) E_1(\mu) \, ,$$
where $E_1(x)$ is the exponential integral.

The FT97 correlation density functional \cite{FT97a,FT97b} is
\begin{equation}
\label{eq:FT97}
\epsilon_c^{FT97} = 0.015545 \left\{f(\mu)
                + \frac{6+4\mu^{1/2}+4\mu}{3+6\mu^{1/2}+6\mu}
                   [\mu f(\mu)+1] \right\} \, .
\end{equation}

In the weak correlation limit
$E_c^{FT97}({\rm He}) \approx -0.05648 +\frac{0.0198}{Z}$, and
$E_c^{FT97}({\rm Hooke}) \approx -0.07076+\frac{0.07}{k^{1/4}}$.
                
\subsection{The One-parameter progressive CS-type functional \cite{Tsuneda}}

This correlation energy density is a variant of the Cole-Salvetti formula,
discussed in subsection 5.1. 
Defining $x_{\sigma}=2^{1/3} \rho^{-4/3} \frac{d\rho}{dr}$,
this functional can be written in the form
\begin{equation}
\label{eq:OP}
\epsilon_c^{OP}=-0.38035 \; \rho
  \frac{\beta_{\alpha\beta}+0.3789}
     {\beta_{\alpha\beta}^4+1.1284\beta_{\alpha\beta}^3
                  +0.3183\beta_{\alpha\beta}^2}
\end{equation}
where
$$\beta_{\alpha\beta}= 0.93934 \rho^{1/3}
  \left[ 3\left(\frac{3}{4\pi}\right)^{1/3} + \frac{0.0084x_{\sigma}^2}
   {1+ 0.0252 x_{\sigma}\log(x_{\sigma}+\sqrt{x_{\sigma}^2+1})}
                        \right] \; .$$

In the weak correlation limit
$E_c^{OP}({\rm He}) \approx -0.08511 + \frac{0.085}{Z}$, and
$E_c^{OP}({\rm Hooke}) \approx -0.09504 + \frac{0.074}{k^{1/4}}$.
                
\subsection{Perdew, Kurth, Zupan and Blaha functional \cite{PKZB}}

This functional depends on the kinetic energy density, and is therefore 
a meta-GGA functional. However, for the ground state
of the two-electron system it reduces to
\begin{equation}
\label{eq:PKZB}
\epsilon_c^{PKZB} = 1.53(\epsilon_c^{PBE} - \epsilon_c^{up}) \, ,
\end{equation}
where $\epsilon_c^{PBE}$ is given by eq. \ref{eq:PBE} and
$$\epsilon_c^{up} = \epsilon_c^{unif,up}+H^{up}(r_s,t) \; .$$
Here $\epsilon_c^{unif, up}$ has the same form as $\epsilon_c^{unif}$,
eq. \ref{eq:LDA}, with the parameters $c_0$, $\alpha_1$ and
${\tilde{\beta}}_i$, $i=1,2,3,4$ being replaced by
\begin{eqnarray*}
{\overline{c}}_0 &=& 0.015545 \\
{\overline{\alpha}}_1 &=& \alpha_1 2^{1/3}=0.25889 \\
{\overline{\beta}}_1 &=& 2{\overline{c}}_0\beta_1^{up}2^{1/6} =0.49273 \\
{\overline{\beta}}_2 &=& 2{\overline{c}}_0\beta_2^{up}2^{1/3} =0.24278 \\
{\overline{\beta}}_3 &=& 2{\overline{c}}_0\beta_3^{up}2^{1/2} =0.14801 \\
{\overline{\beta}}_4 &=& 2{\overline{c}}_0\beta_4^{up}2^{2/3} =0.03085 \; ,
\end{eqnarray*}
respectively. ${\overline{c_0}}$, $\alpha_1$ and $\beta_i^{up}$, $i=1,2,3,4$
are presented in ref. \cite{Perdew92},  Table I.
The fractional powers of 2 take care of the fact that the spin up density is
a half of the total density, so $r_s$ should otherwise have been replaced
by $r_s 2^{1/3}$.

$H^{up}(r_s,t)$ is obtained from $H(r_s,t)$ by multiplying it by
$\frac{1}{2}$, replacing $A$ by
$$A^{up}=\delta\left[\exp\left(-\frac{2\epsilon_c^{unif,up}}{\gamma}\right)
                -1\right]^{-1} \; ,$$
and replacing $t$ by $t^{up}=t\sqrt{2}$.

Inspecting the reparametrized expression we note that
$$\epsilon_c^{unif,up}(r_s) \approx \frac{1}{2}\epsilon_c^{unif}(r_s) \; .$$
This is not apparent in terms of the parameterization given in \cite{Perdew92}.

Numerical integration over the asymptotic density functionals yields
$E_c^{PKZB}({\rm He},Z\rightarrow\infty)=-0.05525$, in agreement with the 
value reported in the erratum to ref. \cite{SSPTD}, and
$E_c^{PKZB}({\rm Hooke},k\rightarrow\infty)=-0.08468$.

\subsection{The TPSS functional \cite{TPSS}}

This functional is of the form
$$\epsilon_c^{TPSS} = \epsilon_c^{PKZB} (1+2.8\epsilon_c^{PKZB})$$
where $\epsilon_c^{PKZB}$ is given by eq. \ref{eq:PKZB}.
The asymptotic correlation energies are
$E_c^{TPSS}({\rm He},Z\rightarrow\infty)=-0.04901$, in agreement with the 
value reported in the erratum to ref. \cite{SSPTD},
 and
$E_c^{TPSS}({\rm Hooke},k\rightarrow\infty)=-0.06843$.

\section{Second-order approximations}

\subsection{Colle-Salvetti formula}

The Colle-Salvetti formula \cite{Colle} for the correlation energy depends on
the Hartree-Fock two-particle density matrix. For the ground state of a 
two-electron atom it can be written in terms of the Hartree-Fock density, providing a correlation energy density
\begin{equation}
\label{eq:CS}
\epsilon_c^{CS} = -0.04918 \; \frac{1
       +0.132\rho^{-5/3}\exp(-0.2533\rho^{-1/3})\; \tau_{CS}}
                          {1+ 0.349\rho^{-1/3}} \, , 
\end{equation}
where
$$\tau_{CS}=\frac{1}{8}\left[\frac{d^2\rho}{dr^2} 
      + \frac{2}{r}\frac{d\rho}{dr} 
      -\frac{1}{\rho}\left(\frac{d\rho}{dr}\right)^2\right] \; .$$
Integration over the asymptotic densities, which can be carried out 
analytically, yields
$E_c^{CS}({\rm He},Z\rightarrow\infty)=-0.01941$
and
$E_c^{CS}({\rm Hooke},k\rightarrow\infty)=-0.001772$, respectively.

\subsection{Lee, Yang and Parr functional \cite{LYP}}

This correlation-energy density is obtained by replacing $\tau_{CS}$,
in eq. \ref{eq:CS}, by
$$\tau_{LYP}=\frac{3}{10}(3\pi^2)^{2/3} \rho^{5/3}
   +\frac{7}{24}\left(\frac{d^2\rho}{dr^2}+\frac{2}{r}\frac{d\rho}{dr}\right)
    -\frac{17}{72\rho}\left(\frac{d\rho}{dr}\right)^2 \; .$$

For the $Z\rightarrow\infty$ (hydrogenic) density 
the LYP correlation energy functional can be integrated analytically, to
yield the correlation energy -0.0566888. This value, obtained in 
ref. \cite{SSPTD}, appears to be too high (in absolute value) to be consistent 
with the values presented in Table 3. Closer examination reveals that the 
asymptotic behavior of the LYP correlation energy for the helium 
isoelectronic sequence is given by
$E_c({\rm He}) \approx -0.0566888+0.009\frac{\log^3(Z)}{Z}$. 
This expression, which is
consistent with the numerical results, can be obtained analytically \cite{Guy}.
It is perhaps worth noting that using the screened hydrogenic density,
in which $Z$ is replaced by $Z-\frac{5}{16}$, the LYP functional yields, for
$Z=2$, the correlation energy -0.04388, compared to the value -0.04378 
obtained with the Hartree-Fock density. 
For Hooke's atom isoelectronic sequence the asymptotic behavior of the 
correlation energy, at $k\rightarrow\infty$, is found to be
$E_c({\rm Hooke}) \approx -0.035501 + 0.00012\frac{\log^3(k)}{k^{1/4}}$.

\subsection{FT98 \cite{FT98}}

This is a modified version of the FT97 functional, extended to include the 
Laplacian of the density. It can be written by replacing $F_{97}$, 
eq. \ref{eq:FF97}, by
$$F_{98}=\frac{1+(0.083726 y_{\sigma})^2}
              {\sqrt{1+0.051581 y_{\sigma}/{r_s}}}
              \exp[-(0.083726 y_{\sigma})^2] \, ,$$
where
$$y_{\sigma}=\frac{1}{(18\pi)^{2/3}} \rho^{-8/3}\left(\frac{d\rho}{dr}\right)^2 
             + \frac{0.89672}{(18\pi)^{4/3}} \rho^{-10/3}
        \left[\frac{1}{\rho} \left(\frac{d\rho}{dr}\right)^2
              - \left(\frac{d^2\rho}{dr^2}
                   +\frac{2}{r}\frac{d\rho}{dr}\right) \right]^2 \, . $$

In the weak correlation limit
$E_c^{FT98}({\rm He}) \approx -0.08173 +\frac{0.17}{Z}$,
and
$E_c^{FT98}({\rm Hooke}) \approx -0.08706 +\frac{0.22}{k^{1/4}}$.

\section{Performance of the various density functionals}

The Hartree-Fock densities were evaluated for both the helium and Hooke's 
isoelectronic sequences, using the procedure described in ref. \cite{Katriel}.
These densities were used to compute the correlation energies in terms of the
various density functionals specified above. The results are 
presented in Tables 1-6, where comparisons with previous pertinent results
are documented, as well as in Figures 1-8. When comparing the different
results it is important to pay attention to the ranges of the correlation
energy scales in the different Figures. Thus, Fig.\ 1, presenting the 0OA
correlation energies for the helium isoelectronic sequence, has a range five 
times as large as Fig.\ 2 (crude 1OA), whose range is three times as large as 
that of Fig.\ 3 (better 1OA). The 2OA results, presented in Fig.\ 4, do 
not improve on the best 1OA results in Fig.\ 3. The overall trend is similar 
for Hooke's
atom isoelectronic sequence, but much less pronounced. Thus, the range of 
Fig.\ 5 (0OA) is twice that of Fig.\ 6 (crude 1OA), where the range of 
deviations of the dft correlation energies from the exact values is only a 
factor of 1.5 broader than that in Fig.\ 7 (better 1OA). Again, the 2OA
correlation energies do not improve on the best 1OA results.

The four 0OA functionals presented in Fig.\ 1 for the helium isoelectronic 
sequence fail to provide an adequate estimate of the correlation
energy. 
Only BR provides the correct 
correlation energy for the helium atom, because it had been calibrated to
do so. Wigner's is the only 0OA-type functional providing a mild dependence 
on the nuclear 
charge ($Z$), that mimics the exact correlation energy, though it is 
more than 
twice as large. In fact, the Wigner and BR correlation-energy functionals 
are the only local density functionals considered, that yield finite 
correlation energies for $Z\rightarrow\infty$. Needless to say, since PW is the
definitive LDA limit, its failings should be taken as a genuine assessment
of this level of approximation. That other 0OA-type correlation energies 
are in closer agreement with the exact results should be interpreted as 
fortuitous. It is regrettable that a ``first-principles'' 
benchmark 1OA correlation energy density cannot be specified in a similar
way, or has not so far been proposed.
For Hooke's atom all zero order approximations considered, presented in 
Fig.\ 5, are roughly equally poor. 
At intermediate values of the force constant the ordering of the various 
approximations relative to the exact correlation energy is similar to that 
for the helium isoelectronic sequence. 

The first-order functionals were split into two classes according 
to their behavior for the helium isoelectronic sequence. The more poorly 
behaved functionals, P86, OP and Becke, are presented in Fig.\ 2. They exhibit 
a strong dependence on the nuclear charge. The five functionals presented 
in Fig.\ 3  all have a reasonable slope, only mildly steeper than that of 
the exact correlation energy. Both PBE and Becke have been parameterized
to yield the exact correlation energy for the helium atom. For Hooke's 
atom isoelectronic sequence most 1OA results, except P86, cluster together,
being roughly twice as large as the exact correlation energy. We note that
TPSS and FT97 are remarkably close to one another; so are PBE and PKZB. Neither pair remains close for the helium isoelectronic sequence. Rather, PBE and TPSS
behave similarly; so do FT97 and PKZB, even more closely but further away 
from the exact correlation energy. These contradictory patterns serve as a
warning against assigning excessive significance to limited scope
examination of the comparative performance of different density functionals.

The three second-order approximations considered are presented, for the
helium isoelectronic sequence, in Fig.\ 4. LYP has the overall most satisfactory 
behavior, although for $Z<6$ CS is in closer agreement with the exact
correlation energies. For Hooke's sequence ({\it {cf.}} Fig.\ 8), none
of the second-order approximations provides a good agreement with the exact 
correlation energies. LYP and CS, which are conceptually related, are 
remarkably close to one another, along Hooke's atom isoelectronic sequence; 
however, this is not true for the helium isoelectronic sequence.

As already pointed out, Hooke's atom isoelectronic sequence allows the 
examination of the weak binding (high correlation) limit, $k\rightarrow 0$.
The computed correlation energies for the range of lower force constants 
suggest that
$$\frac{E_c^{PW}}{E_c^{exact}} \approx 1 +1.1k^{\frac{1}{16}} \, ,$$
and 
$$\frac{E_c^{GL}}{E_c^{exact}} \approx 1 + 2.18k^{\frac{1}{8}} \, ,$$
although, in both cases, the correlation energies are considerably higher 
than the exact correlation energies, even for the lowest force constants 
considered.
These results may be taken to suggest that at very low force constants the 
local density approximation becomes exact, since the density is spread out and its 
gradient becomes negligibly low. A similar behavior is observed for
the ratios $\frac{E_c^{FT97}}{E_c^{exact}}$,
$\frac{E_c^{PKZB}}{E_c^{exact}}$ and $\frac{E_c^{TPSS}}{E_c^{exact}}$.
This feature has not been deliberately built into any of these approximations.
It is, therefore, noteworthy that it holds to the extent that it does.

Our study clearly suggests that the approximate density functionals
in general perform better for the helium sequence than for Hooke's atoms.
In order to understand this behavior we show in Fig.\ 9 scaled electron
densities, i.e., $\rho/c^3$, as well as $\frac{d\rho}{dr}/c^4$ and 
$\frac{d^2\rho}{dr^2}/c^5$, as functions of $x=c\cdot r$ with $c$ being 
$Z-5/16$ for the He sequence and $k^{1/4}-0.06649$ for Hooke's atoms.
These scaling parameters were derived at the bottom of section 2, above. 

A number of observations can be made from the figure. First, the curves for
the He sequence are all remarkably similar, except for the region $r\to 0$, 
where
the cusp problem becomes visible, in particular for the second-order derivative.
An analogous similarity is not found for Hooke's atoms. The curves 
corresponding to the highest values of $k$ are indeed very close to the
expected $k\rightarrow\infty$ limit, but the range of $k$ is broad enough
to allow significant deviations from this limiting behavior. 
It follows that the He sequence does not provide a useful
testing ground for exploring approximate density functionals, as long as it 
is not extended with further test systems. 

Second, the electron density is more confined in space for the He sequence than
for Hooke's atoms. For the former, the cusp problem may be important since 
the region for $r\to0$ is one of a large energy density although its volume is
vanishing. A similar region does not exist for Hooke's atoms, i.e., for 
those systems the electron density is not maximized in a region of 
diverging
potential. This suggests a further reason for using Hooke's atoms as model 
systems
for studying properties of electron densities in regions where the potential 
stays finite. This is exactly what is relevant when developing approximate 
density functionals aimed at describing the properties of chemical bonds.

\section{Conclusions}

A word is appropriate about the choice of density functionals investigated.
We did try to be fairly exhaustive. Some functionals were included
primarily to respect the pioneers, most notably Wigner, although we have 
not made justice to contributions by Pines \cite{Pines}, 
Gell-Mann and Brueckner \cite{GMB}, Hedin \cite{Hedin}, and many others. 
We did exclude 
a couple of recent functionals either because the correlation energies
they provide for helium were judged to be poor, because no result was 
available to verify our implementation, or because they were considerably
harder to implement. In the cases excluded we did not have a compelling reason
to expect a particularly good performance for the helium sequence, let alone
Hooke's sequence. We may have excluded highly worth-while functionals 
by oversight or ignorance, for which we apologize.

It has already been noted by several authors that most density functionals
fail to yield an adequate dependence of the correlation energy on the
nuclear charge, $Z$, along isoelectronic sequences. However, some, such as 
PBE, TPSS, CS and LYP are not too bad. None of that is true for Hooke's 
sequence, suggesting that the ultimate aim of developing a universal density 
functional has not yet been attained, not even for spherically-symmetric
two-fermion systems. By analysing the electron densities for the two
classes of systems, we found that the He sequence does not provide a 
sufficiently general class of systems for assessing or improving 
the quality of density functionals, whereas this is to a larger extent
the case for Hooke's atom isoelectronic sequence. A feasible approach to
improving the universality of correlation energy density functionals could 
be to examine their reparametrization, allowing for a dependence on the 
one-body potential. 
Since this potential is uniquely determined by the 
ground-state Hartree-Fock density, this approach is still consistent with the 
Hohenberg-Kohn framework of dft.

\noindent
{\bf Acknowledgements:} Helpful correspondence with Dr. V. N. Staroverov 
and with Professor E. R. Davidson is gratefully acknowledged. 
So is a helpful discussion with Dr. G. Katriel. 
This work was supported by the German Research Council through the SFB 277
at the University of Saarland.

\vfill\eject

\newpage

\leavevmode
\begin{table}
\begin{center}
\begin{tabular}{|l|l|l|l|l|l|}
\hline
  Z &  $\;\; E_c^{exact}$   &  $\;\; E_c^{PW}$ &  $\;\; E_c^{GL}$ &  
     $\;\; E_c^{BR}$ & $\;\; E_c^{W}$ \\
\hline
  2 &  -0.042044 & -0.1125[a]  &    -0.1385 &-0.04204 & -0.0959 \\
\hline
  3 &  -0.043498 & -0.1346[b]  &    -0.1666 & -0.05922 & -0.1015 \\
\hline
  4 &  -0.044267 & -0.1504[a]  &    -0.1862 &-0.07276 &-0.1043 \\
\hline
  5 &  -0.044736 & -0.1628  &    -0.2037 &-0.08376 &-0.1059 \\
\hline
  6 &  -0.045054 & -0.1730  &    -0.2137 &-0.09290 &-0.1071 \\
\hline
  7 &  -0.045281 & -0.1817  &    -0.2240 &-0.10065 &-0.1079\\
\hline
  8 &  -0.045452 & -0.1893  &    -0.2330 &-0.10730 &-0.1085\\
\hline
  9 &  -0.045586 & -0.1960  &    -0.2409 &-0.11309 &-0.1090 \\
\hline
 10 & -0.045692  & -0.2020[a]  &    -0.2499 &-0.11819 &-0.1093 \\
\hline
 20 &  $\>$      & -0.2424  &    -0.2944 &-0.14834 &-0.1110 \\
\hline
 50 &   $\>$     & -0.2971  &    -0.3555 &-0.17618 &-0.1121 \\
\hline
\end{tabular}
\end{center}

\noindent
\hspace{2.8cm} [a] ref. \cite{UG}. [b] ref. \cite{PT2004}.
\caption{Zero-order-approximation correlation energies for the Helium isoelectronic sequence. The approximate
density functionals are defined in the text.}
\label{tab01}
\end{table}

\begin{sidewaystable}
\begin{center}
\begin{tabular}{|l|l|l|l|l|l|l|l|l|l|}
\hline
  Z &  $\;\; E_c^{exact}$  & $\;\; E_c^{PBE}$  &  $\;\; E_c^{PKZB}$ & 
   $\;\; E_c^{OP}$ & $\;\; E_c^{FT97}$ & $\;\; E_c^{TPSS}$  &  $\;\; E_c^{B88}$ & 
      $\;\; E_c^{P86}$  &  $\;\; E_c^{WL}$ \\
\hline
  2  & -0.042044 & -0.04202[b] &  -0.04727[b] & -0.05178[c] & -0.04623[d] &
           -0.04306[b] & -0.04189[e] & -0.04390[f]  & -0.04204[a] \\
\hline
  3  & -0.043498 & -0.04478[b] &  -0.05063[b] & -0.06089 & -0.04952 &
           -0.04574[b] & -0.04985[e] & -0.04542[f]  & -0.04418[e] \\
\hline
  4  & -0.044267 & -0.04606[b] &  -0.05224[b] & -0.06607 & -0.05119 &
           -0.04698[b] & -0.05462[e] & -0.04868[f]  & -0.04520[e] \\
\hline
  5  & -0.044736 & -0.04676 &  -0.05316 & -0.06943 & -0.05222 &
           -0.04767 & -0.05782 & -0.05324  & -0.04579 \\
\hline
  6  & -0.045054 & -0.04720 &  -0.05374 & -0.07178 & -0.05288 &
           -0.04811 & -0.06012 & -0.05865  & -0.04618 \\
\hline
  7  & -0.045281 & -0.04748 &  -0.05441 & -0.07351 & -0.05338 &
           -0.04839 & -0.06186 & -0.06457  & -0.04645 \\
\hline
  8  & -0.045452 & -0.04768 &  -0.05441 & -0.07485 & -0.05375 &
           -0.04860 & -0.06322 & -0.07079  & -0.04665 \\
\hline
  9  & -0.045586 & -0.04782 &  -0.05463 & -0.07591 & -0.05404 &
           -0.04874 & -0.06431 & -0.07717  & -0.04681 \\
\hline
 10  & -0.045692 & -0.04793 &  -0.05478 & -0.07677 & -0.05428 &
           -0.04885 & -0.06521 & -0.08360  & -0.04693[a] \\
\hline
 20  &            & -0.04826 &  -0.05536 & -0.08080 & -0.05535 &
       -0.04923 & -0.0695 & -0.14274 & -0.04749 \\
\hline
50   &           & -0.04824 &  -0.05550 & -0.08335 & -0.05602 &
           -0.04928 & -0.07240 & -0.24874  & -0.04781 \\
\hline
\end{tabular}
\end{center}

\noindent
\hspace{0.2cm} 
[a] ref. \cite{UG}. [b] ref. \cite{PT2004}. [c] ref. \cite{Tsuneda}.
[d] ref. \cite{FT98}. [e] ref. \cite{WIv}. [f] ref. \cite{P86}.
\caption{First-order-approximation 
correlation energies for the Helium isoelectronic sequence. The approximate
density functionals are defined in the text.}
\label{tab02}
\end{sidewaystable}

\newpage

\leavevmode
\begin{table}
\begin{center}
\begin{tabular}{|l|l|l|l|l|}
\hline
 Z & $\;\; E_c^{exact}$ & $\;\; E_c^{CS}$ & $\;\; E_c^{LYP}$ & $\;\; E_c^{FT98}$  \\\hline
  2 & -0.042044 & -0.04158[c] & -0.04378[a] &   -0.04643[d] \\
\hline
  3 & -0.043498 & -0.0439 [g] & -0.04755[e] &   -0.05254 \\
\hline
  4 & -0.044267 & -0.04421[g] & -0.04905[a] &   -0.05804 \\
\hline
  5 & -0.044736 & -0.04385 & -0.04972 &   -0.06122 \\
\hline
  6 & -0.045054 & -0.04326 & -0.05004 &   -0.06360 \\
\hline
  7 & -0.045281 & -0.04260 & -0.05018 &   -0.06546 \\
\hline
  8 & -0.045452 & -0.04191 & -0.05022 &   -0.06697 \\
\hline
  9 & -0.045586 & -0.04124 & -0.05023 &   -0.06821 \\
\hline
 10 & -0.045692 & -0.04060 & -0.05021[a] &   -0.06925 \\
\hline
 20 & $\>$      & -0.03591 & -0.04992 &   -0.07464 \\
\hline
 50 & $\>$      & -0.03001 & -0.05008 &   -0.07862 \\
\hline
\end{tabular}
\end{center}

\noindent
\hspace{3.1cm}
[a] ref. \cite{UG}. [c] ref. \cite{Tsuneda}. [d] ref. \cite{FT98} 
[e] ref. \cite{WIv}.  [g] ref. \cite{LYP}. 
\caption{Second-order-approximation 
correlation energies for the Helium isoelectronic sequence. The approximate
density functionals are defined in the text.}
\label{tab03}
\end{table}

\newpage

\leavevmode
\begin{table}
\begin{center}
\begin{tabular}{|l|l|l|l|l|l|}
\hline
  $\;\;\; k$ & $\;\; E_c^{exact}$ & $\;\; E_c^{PW}$ & $\;\; E_c^{GL}$  & 
    $\;\; E_c^{BR}$  & $\;\; E_c^W$ \\
\hline
 0.000300891& -0.01691& -0.02865 &  -0.0303 &-0.00392& -0.0388\\
\hline
 0.00133497 & -0.02198& -0.03857 &  -0.0425 &-0.00609& -0.0506\\
\hline
 0.01       & -0.02904& -0.05478[h] &  -0.0633 &-0.01071& -0.0665\\
\hline
 0.25       & -0.03843& -0.08614[i] &  -0.1045 &-0.02424& -0.0871\\
\hline
 1.0        & -0.04134& -0.1013 [h]  &  -0.1243 &-0.03321& -0.0935\\
\hline
16.        & -0.04527& -0.1342  &  -0.1663 &-0.05784& -0.1023\\
\hline
100.       & -0.04684& -0.1575 [h]  &  -0.1952 &-0.07847& -0.1060\\
\hline
10000.       & -0.04877& -0.2208 [h]  &  -0.2699 &-0.13417& -0.1105\\
\hline
\end{tabular}
\end{center}

\noindent
\hspace{2cm} [h] ref. \cite{HU}. [i] ref. \cite{BPE97}

\caption{Zero-order-approximation 
correlation energies for Hooke's atom isoelectronic sequence. The approximate
density functionals are defined in the text.}
\label{tab04}
\end{table}

\leavevmode
\begin{sidewaystable}
\begin{center}
\begin{tabular}{|l|l|l|l|l|l|l|l|l|l|}
\hline
 $\;\;\; k$ & $\;\; E_c^{exact}$ & $\;\; E_c^{PBE}$ & $\;\; E_c^{PKZB}$ & 
   $\;\; E_c^{OP}$ & $\;\; E_c^{FT97}$ & $\;\; E_c^{TPSS}$
   &  $\;\; E_c^{B88}$ & $\;\; E_c^{P86}$ &  $\;\; E_c^{WL}$  \\
\hline
 0.000300891 & -0.01691 & -0.02148 & -0.01895 & -0.00970 & -0.01904
         & -0.01835 & -0.00860 & -0.02508 & -0.0385   \\
\hline
 0.00133497  & -0.02198 & -0.02782 & -0.02525 & -0.01446 & -0.02483
         & -0.02416 & -0.01247 & -0.03244 & -0.0475   \\
\hline
 0.01        & -0.02904 & -0.03705 & -0.03470 & -0.02341 & -0.03292
         & -0.03260 & -0.01951 & -0.04216 & -0.0577   \\
\hline
 0.25        & -0.03843 & -0.05118[j] & -0.04968 & -0.04265 & -0.04484
         & -0.04523 & -0.03448 & -0.05434 & -0.0684   \\
\hline
 1.0         & -0.04134 & -0.05748 & -0.05546 & -0.05159 & -0.04924
         & -0.04981 & -0.04168 & -0.05833 & -0.0712   \\
\hline
16.         & -0.04527 & -0.06520 & -0.06522 & -0.06768 & -0.05652
         & -0.05708 & -0.05547 & -0.06712 & -0.0746   \\
\hline
100.        & -0.04684 & -0.06960 & -0.07028 & -0.07594 & -0.06023
         & -0.06057 & -0.06315 & -0.07683 & -0.0759   \\
\hline
10000.        & -0.04877 & -0.07659[k] & -0.07861 & -0.08816 & -0.06624
         & -0.06569 & -0.07563 & -0.12609 & -0.0774   \\
\hline
\end{tabular}
\end{center}

\noindent
\hspace{0.6cm}
[j] ref. \cite{LCB}. [k] ref. \cite{BCL}.

\caption{First-order-approximation 
correlation energies for Hooke's atom isoelectronic sequence. The approximate
density functionals are defined in the text.}
\label{tab05}
\end{sidewaystable}

\newpage

\leavevmode
\begin{table}
\begin{center}
\begin{tabular}{|l|l|l|l|l|}
\hline
  $\;\;\; k$ & $\;\; E_c^{exact}$  & $\;\; E_c^{CS}$ & $\;\; E_c^{LYP}$ & 
  $\;\; E_c^{FT98}$  \\
\hline
 0.000300891& -0.01691 & -0.01068 & -0.01069 &  -0.01141 \\
\hline
 0.00133497 & -0.02198 & -0.01559 & -0.01561 &  -0.01563 \\
\hline
 0.01       & -0.02904 & -0.02367 & -0.02369 &  -0.02239 \\
\hline
 0.25       & -0.03843 & -0.03542 & -0.03502[m] &  -0.03488 \\
\hline
 1.0        & -0.04134 & -0.03837 & -0.03762 &  -0.04057 \\
\hline
16.        & -0.04527 & -0.03913 & -0.03799 &  -0.05183 \\
\hline
100.       & -0.04684 & -0.03636 & -0.03573 &  -0.05879 \\
\hline
10000.       & -0.04877 & -0.02397 & -0.02871[m] &  -0.07296 \\
\hline
\end{tabular}
\end{center}

\noindent
\hspace{2.8cm} [m] ref. \cite{FUT}.
\caption{Second-order-approximation 
correlation energies for Hooke's atom isoelectronic sequence. The approximate
density functionals are defined in the text.}
\label{tab06}
\end{table}

\newpage

\unitlength1cm
\begin{figure}
\begin{center}
\begin{picture}(15,10)
\put(0,0){\psfig{file=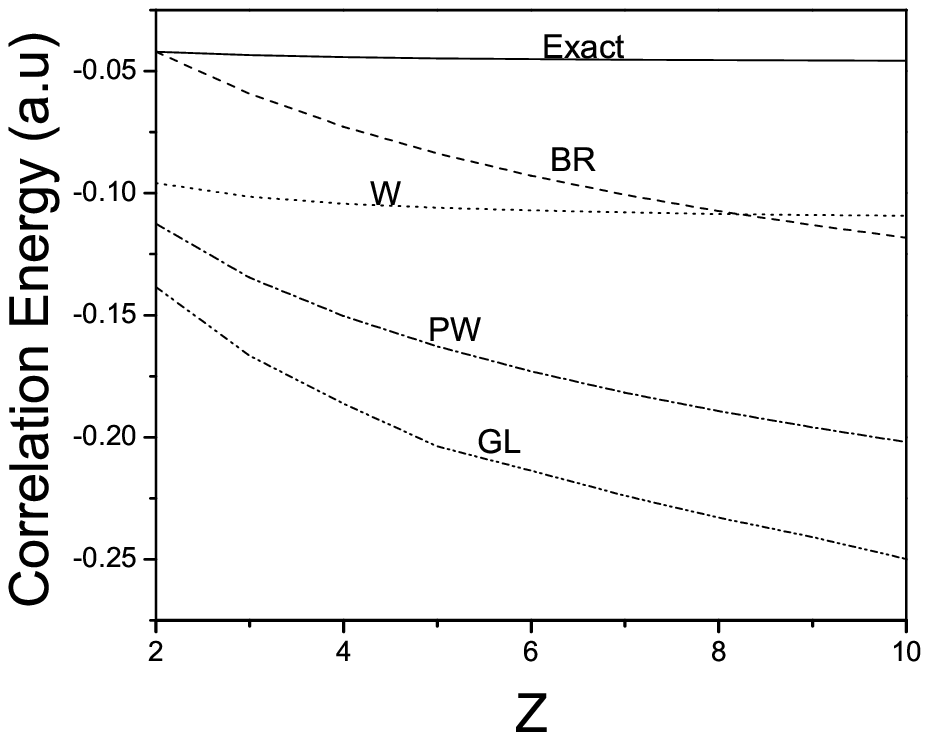,width=15cm}}
\end{picture}
\caption{Zero-order-approximation 
correlation energies for the helium isoelectronic sequence. The approximate
density functionals are defined in the text.}
\label{fig01}
\end{center}
\end{figure}

\unitlength1cm
\begin{figure}
\begin{center}
\begin{picture}(15,10)
\put(0,0){\psfig{file=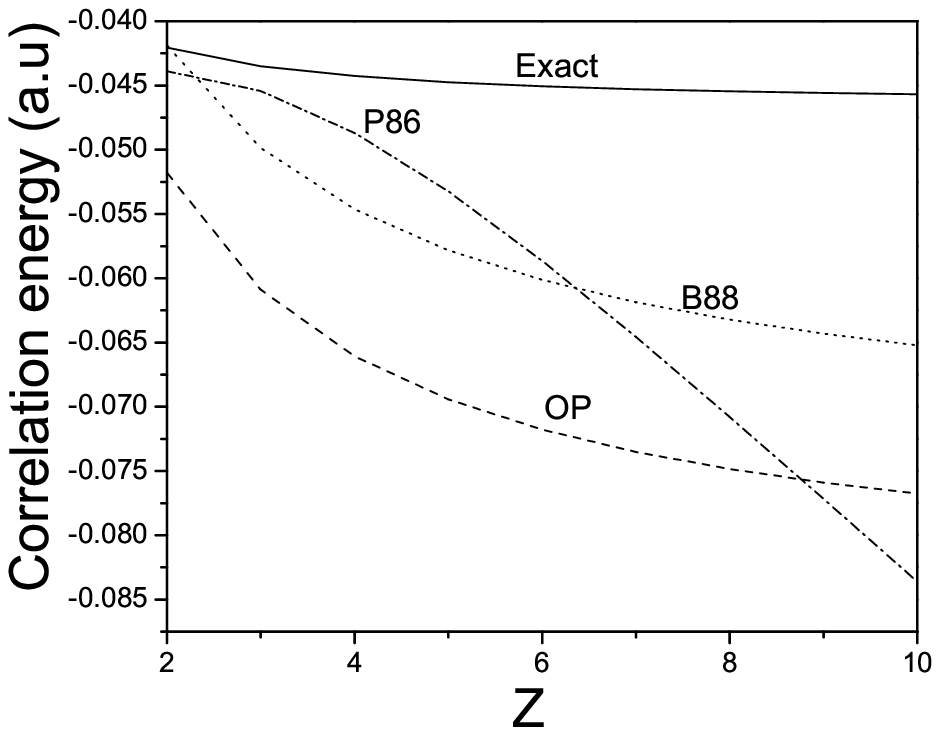,width=15cm}}
\end{picture}
\caption{First-order-approximation 
correlation energies for the helium isoelectronic sequence for less-well performing
functionals. The approximate
density functionals are defined in the text.}
\label{fig02}
\end{center}
\end{figure}

\unitlength1cm
\begin{figure}
\begin{center}
\begin{picture}(15,10)
\put(0,0){\psfig{file=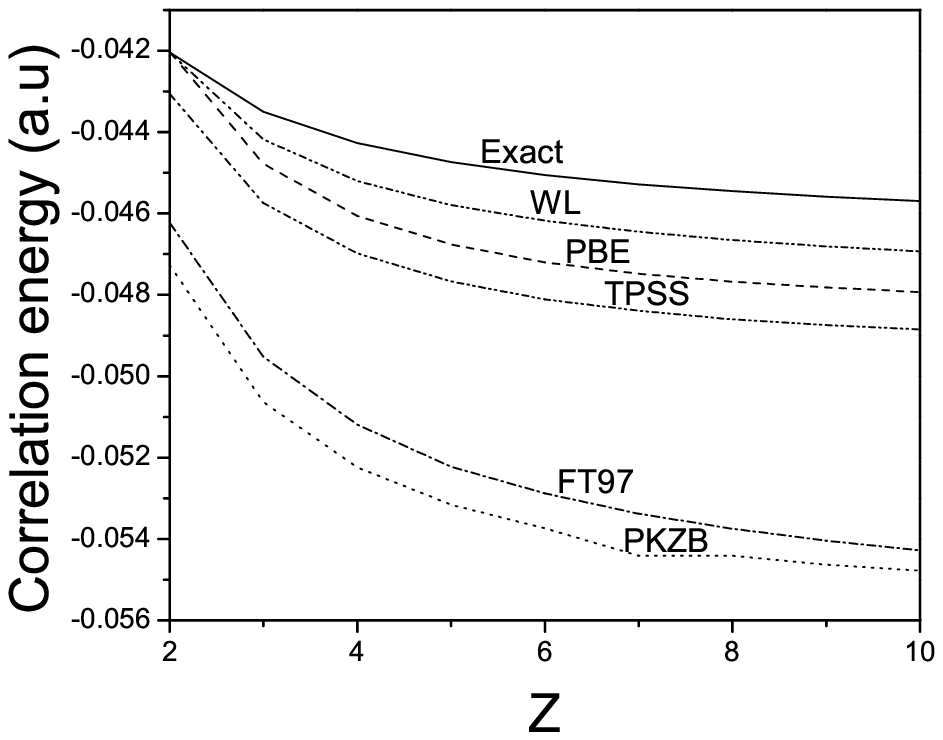,width=15cm}}
\end{picture}
\caption{First-order-approximation 
correlation energies for the helium isoelectronic sequence for well performing
functionals. The approximate
density functionals are defined in the text.}
\label{fig03}
\end{center}
\end{figure}

\unitlength1cm
\begin{figure}
\begin{center}
\begin{picture}(15,10)
\put(0,0){\psfig{file=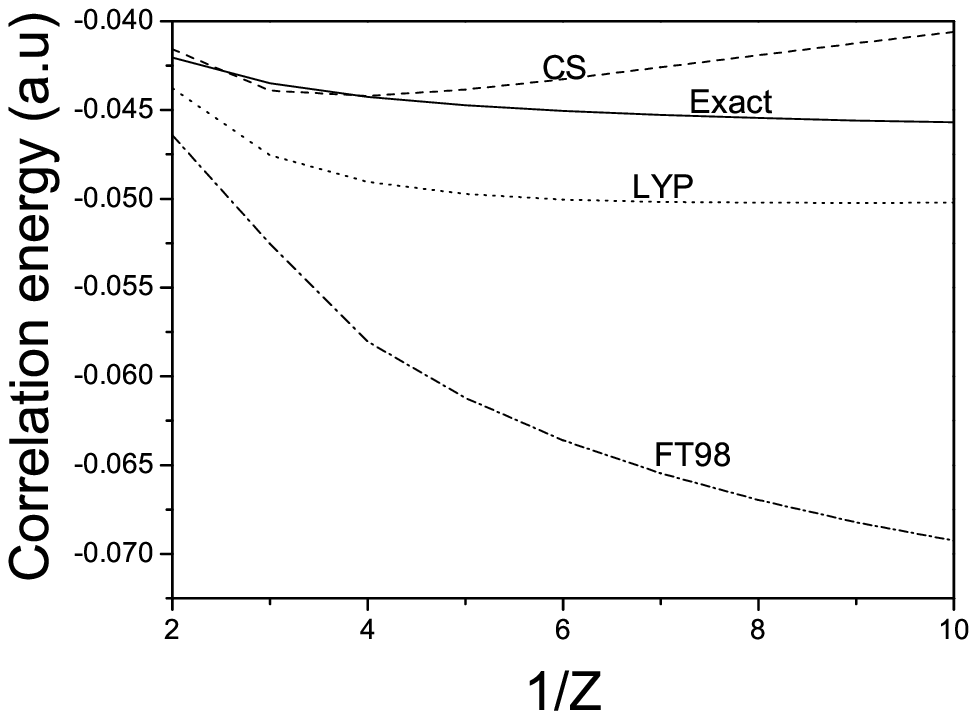,width=15cm}}
\end{picture}
\caption{Second-order-approximation 
correlation energies for the helium isoelectronic sequence. The approximate
density functionals are defined in the text.}
\label{fig04}
\end{center}
\end{figure}

\unitlength1cm
\begin{figure}
\begin{center}
\begin{picture}(15,10)
\put(0,0){\psfig{file=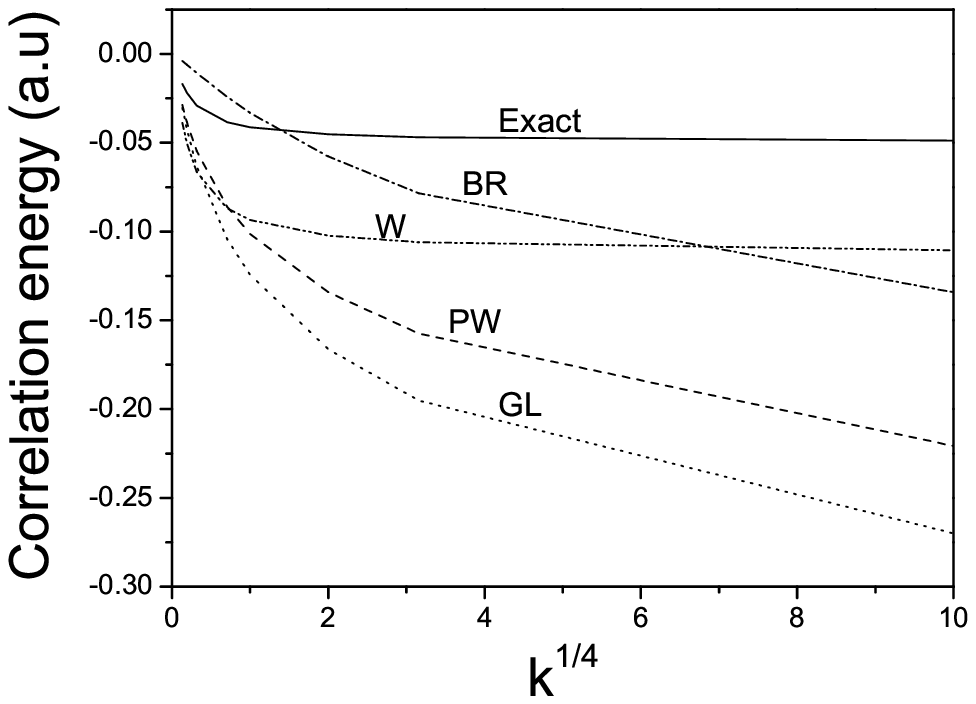,width=15cm}}
\end{picture}
\caption{Zero-order-approximation 
correlation energies for Hooke's isoelectronic sequence. The approximate
density functionals are defined in the text.}
\label{fig05}
\end{center}
\end{figure}

\unitlength1cm
\begin{figure}
\begin{center}
\begin{picture}(15,10)
\put(0,0){\psfig{file=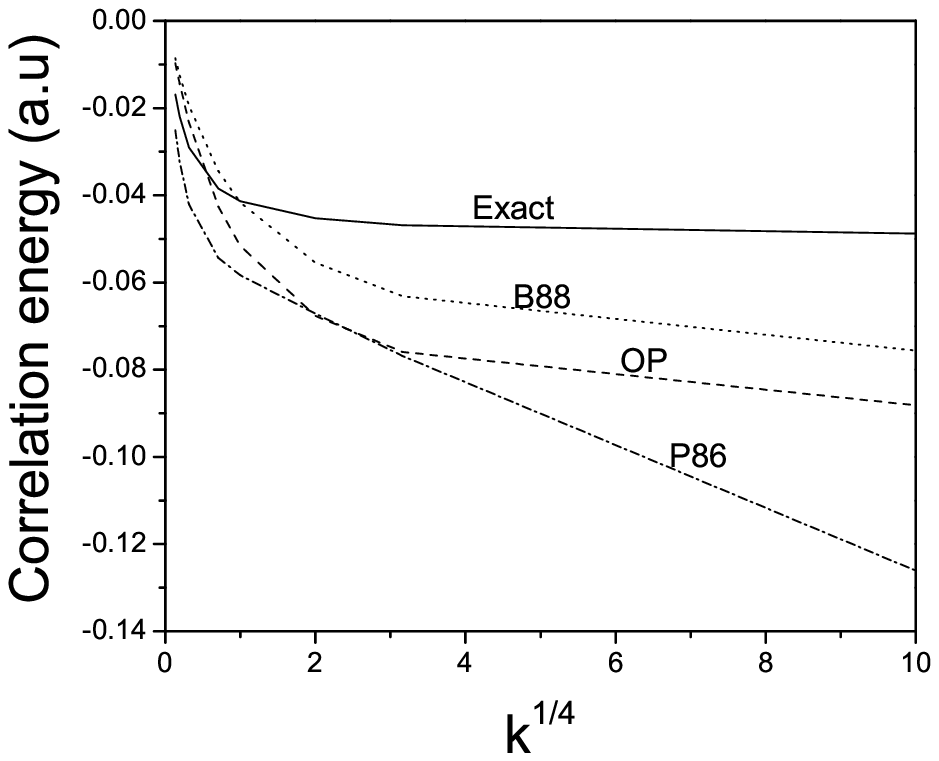,width=15cm}}
\end{picture}
\caption{First-order-approximation 
correlation energies for Hooke's isoelectronic sequence for less-well performing
functionals. The approximate
density functionals are defined in the text.}
\label{fig06}
\end{center}
\end{figure}

\unitlength1cm
\begin{figure}
\begin{center}
\begin{picture}(15,10)
\put(0,0){\psfig{file=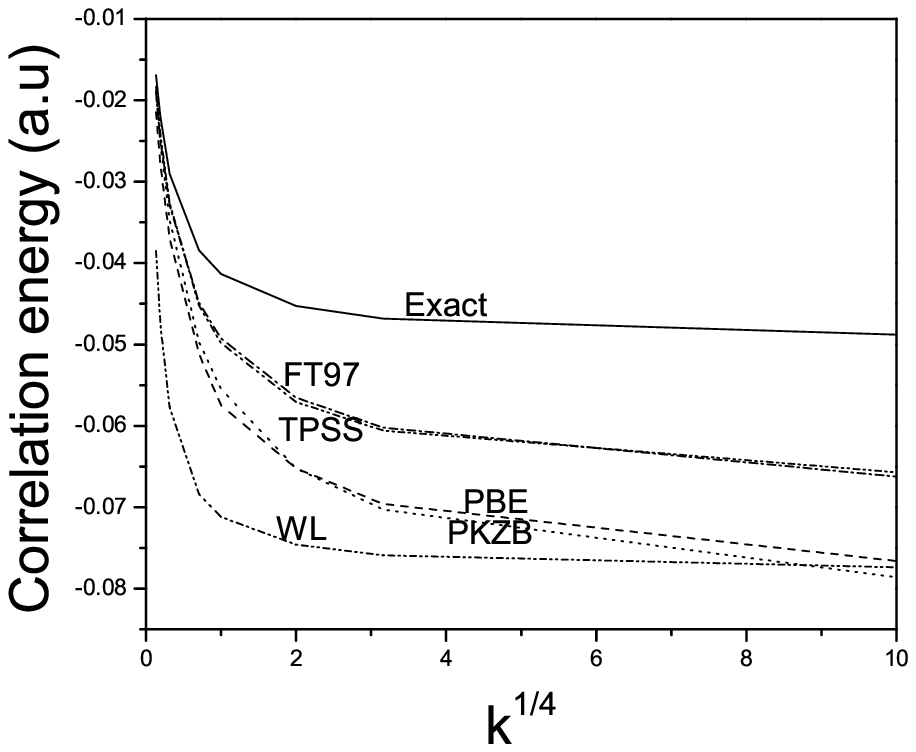,width=15cm}}
\end{picture}
\caption{First-order-approximation 
correlation energies for Hooke's isoelectronic sequence for well performing
functionals. The approximate
density functionals are defined in the text.}
\label{fig07}
\end{center}
\end{figure}

\unitlength1cm
\begin{figure}
\begin{center}
\begin{picture}(15,10)
\put(0,0){\psfig{file=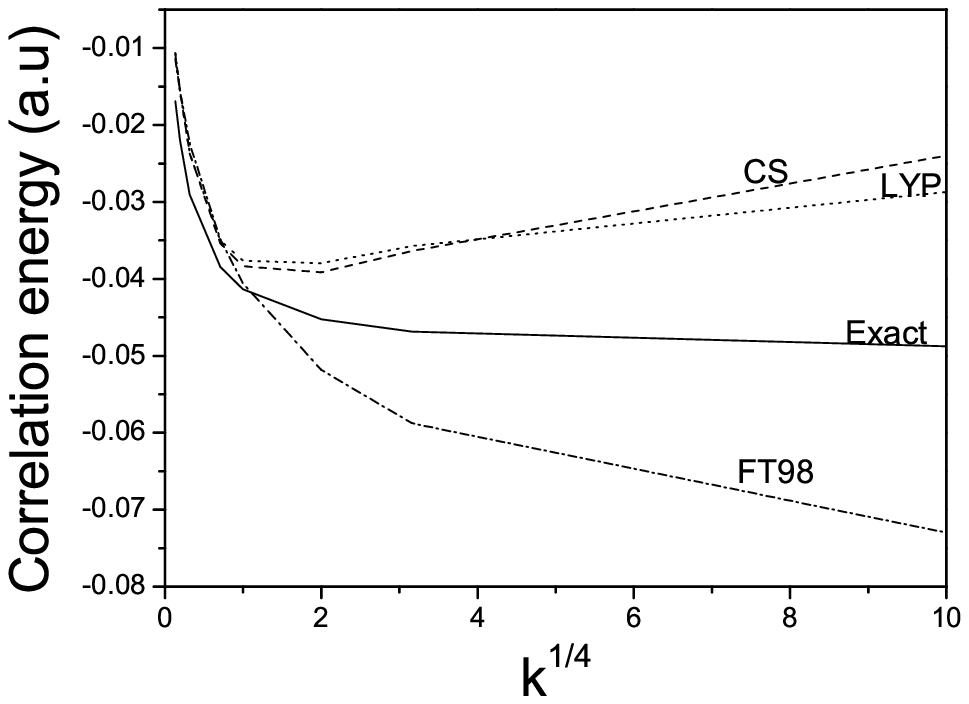,width=15cm}}
\end{picture}
\caption{Second-order-approximation 
correlation energies for Hooke's isoelectronic sequence. The approximate
density functionals are defined in the text.}
\label{fig08}
\end{center}
\end{figure}

\unitlength1cm
\begin{figure}
\begin{center}
\begin{picture}(10,15)
\put(0,0){\psfig{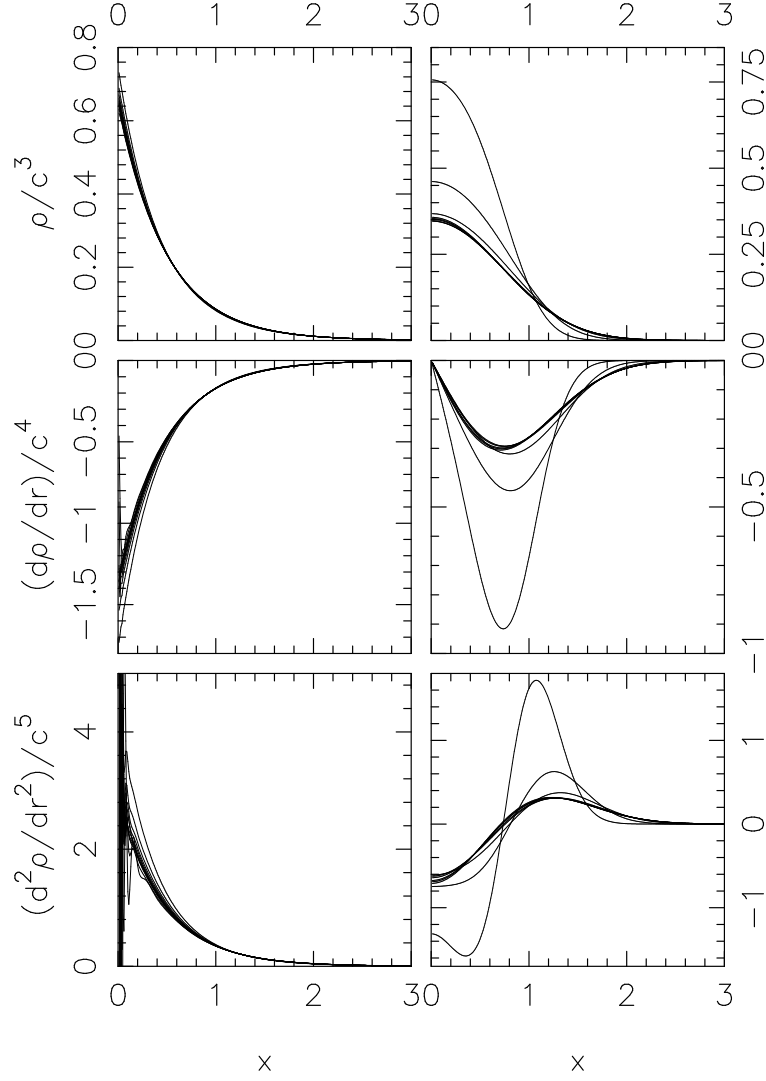}}
\end{picture}
\caption{The electron density and its derivatives for the helium
isoelectronic sequence (left panels) and Hooke's atom isoelectronic 
sequence (right panels). 
All curves are given as functions of a scaled radial coordinate $x=c\cdot r$,
with $c$ being $Z-5/16$ for 
the He sequence and $k^{1/4}-0.06649$ for Hooke's atoms. For Hooke's atoms, 
the uppermost curves for small $x$ in the upper panel and the lowest curves 
for small $x$ in the other two panels correspond to the largest values of $k$.} 
\label{fig09} 
\end{center}
\end{figure}


\begin{thebibliography}{99}

\bibitem{Kurth} S. Kurth, J. P. Perdew and P. Blaha,
    Int. J. Quant. Chem. {\bf 75}, 889 (1999).

\bibitem{Chak} S. J. Chakravorty, S. R. Gwaltney, E. R. Davidson,
    F. A. Parpia and C. Froese Fischer,
    Phys. Rev. A {\bf 47}, 3649 (1993).

\bibitem{Jarz} A. A. Jarzecki and E. R. Davidson,
             Phys. Rev. A {\bf 58}, 1902 (1998).

\bibitem{LYP} C. Lee, W. Yang and R. G. Parr, Phys. Rev. B {\bf 37}, 785 (1988)

\bibitem{Colle} R. Colle and O. Salvetti,
              Theor. Chim.  Acta {\bf 37}, 329 (1975).

\bibitem{SPP} H. Stoll, C. M. E. Pavlidou and H. Preuss,
       Theor. Chim. Acta {\bf 49},143 (1978).

\bibitem{PZ} J. P. Perdew and A. Zunger, Phys. Rev. B {\bf 23}, 5048 (1981).

\bibitem{Be} A. D. Becke, J. Chem. Phys. {\bf 88}, 1053 (1988).

\bibitem{LC} G. C. Lie and E. Clementi, J. Chem. Phys. {\bf 60}, 1275 (1974).

\bibitem{P86} J. P. Perdew, Phys. Rev. B {\bf 33}, 8822 (1986).

\bibitem{Perdew92} J. P. Perdew and Y. Wang,
              Phys. Rev. B {\bf 45}, 13244 (1992).

\bibitem{UG} C. J. Umrigar and X. Gonze, Phys. Rev. A {\bf 50}, 3827 (1994).

\bibitem{WB} T. K. Whittingham and K. Burke,
             J. Chem . Phys. {\bf 122}, 134108 (2005).

\bibitem{PBE} J. P. Perdew, K. Burke and M. Ernzerhof,
              Phys. Rev. Lett. {\bf 77}, 3865 (1996); {\it ibid} {\bf 78}, 1396 (1997)
              (errata).

\bibitem{Tong} B. Y. Tong, Phys Rev. A {\bf 4}, 1375 (1971).

\bibitem{RS} J. Rey and A. Savin, Int. J. Quantum Chem. {\bf 69}, 581 (1998).

\bibitem{SSPTD} V. N. Staroverov, G. E. Scuseria, J. P. Perdew, J. Tao and
   E. R. Davidson, Phys. Rev. A {\bf 70}, 012502 (2004). erratum (forthcoming).

\bibitem{PRTSSC} J. P. Perdew, A. Ruzsinszky, J. Tao, V. N. Staroverov,
   G. E. Scuseria and G. I. Csonka, J. Chem. Phys. {\bf 123}, 062201 (2005).

\bibitem{TPSS} J. Tao, J. P. Perdew, V. N. Staroverov and G. E. Scuseria,
           Phys. Rev. Lett. {\bf 91}, 146401 (2003).

\bibitem{Kestner} N. R. Kestner and O. Sinanoglu,
                      Phys. Rev. {\bf 128}, 2687 (1962).

\bibitem{KaisandLevine} S. Kais, D. R. Herschbach and R. D. Levine, 
            J. Chem. Phys. {\bf 91}, 7791 (1989 ).

\bibitem{Taut} M. Taut, Phys. Rev. A {\bf 48}, 3561 (1993).

\bibitem{Kais} S. Kais, D. R. Herschbach, N. C. Handy, C. W. Murray
        and G. J. Laming, J. Chem. Phys. {\bf 99}, 417 (1993).

\bibitem{Laufer} P. M. Laufer and J. B. Krieger,
                      Phys. Rev. A {\bf 33}, 1480 (1986).

\bibitem{Henderson} T. M. Henderson and R. J. Bartlett,
                    Phys. Rev. A {\bf 70}, 022512 (2004).  

\bibitem{Burke} K. Burke, J. P. Perdew and D. C. Langreth, 
                 Phys. Rev. Lett. {\bf 73}, 1283 (1994).

\bibitem{BPE97} K. Burke, J. P. Perdew and M. Ernzerhof,
                 Int. J. Quantum Chem. {\bf 61}, 287 (1997).

\bibitem{LCB} K.-C. Lam, F. G. Cruz and K. Burke,
                 Int. J. Quantum Chem. {\bf 69}, 533 (1998).

\bibitem{BCL} K. Burke, F. G. Cruz and K.-C. Lam,
                 Int. J. Quantum Chem. {\bf 70}, 583 (1998).

\bibitem{IBL} S. Ivanov, K. Burke, and M. Levy, 
                 J. Chem. Phys. {\bf 110}, 10262 (1999).

\bibitem{Ludena} A. Artemyev, E.V. Ludena, and V. Karasiev,
                 J. Mol. Struct.: THEOCHEM {\bf 580}, 47 (2002);
                 E. V. Ludena, D. Gomez, V. Karasiev and P. Nieto,
                 Int. J. Quantum Chem.  {\bf99}, 297 (2004).

\bibitem {HU} C.-J. Huang and C. J. Umrigar, Phys. Rev. A {\bf 56}, 290 (1997).

\bibitem{JGAG} J. Jung, P. Garc\'\i a-Gonz\'alez, J. E. Alvarellos and R. W. Godby,
                 Phys. Rev. A {\bf 69}, 052501 (2004).

\bibitem{TP} J. Tao and J. P. Perdew in {\underline{Reviews of Modern 
     Quantum Chemistry}}, Vol. I, K. D. Sen, Ed., 
     World Scientific, New Jersey 2002. p. 719.

\bibitem{PKZB} J. P. Perdew, S. Kurth, A. Zupan and P. Blaha,
               Phys. Rev. Lett. {\bf 82}, 2544 (1999).

\bibitem{Brown} R. J. White and W. Byers Brown, 
                   J. Chem. Phys. {\bf 53}, 3869 (1970).

\bibitem{Wigner34} E. Wigner, Phys. Rev. {\bf 46}, 1002 (1934).

\bibitem{Wigner38} E. P. Wigner, Trans. Farady Soc. {\bf 34}, 678 (1938).

\bibitem{GL} O. Gunnarsson and B. I. Lundqvist,
           Phys. Rev. B {\bf 13}, 4274 (1976).

\bibitem{McW} R. McWeeny, Nature {\bf 166}, 21 (1950).

\bibitem{BR} G. Brual and S. M. Rothstein, 
                     J. Chem. Phys. {\bf 69}, 1177 (1978).

\bibitem{WL} L. C. Wilson and M. Levy, Phys. Rev. B {\bf 41}, 12930 (1990).


\bibitem{FT97a} M. Filatov and W. Thiel, 
            Int. J. Quant. Chem. {\bf 62}, 603 (1997)

\bibitem{FT97b} M. Filatov and W. Thiel, Molec. Phys. {\bf 91}, 847 (1997). 

\bibitem{FT98} M. Filatov and W. Thiel, Phys. Rev. A {\bf 57}, 189 (1998).

\bibitem{Tsuneda} T. Tsuneda, T. Suzumura and K. Hirao,
    J. Chem. Phys. {\bf 110}, 10664 (1999).

\bibitem{Star} V. Staroverov, personal communication.

\bibitem{Guy} G. Katriel, personal communication.

\bibitem{Katriel}
J. Katriel, S. Roy and M. Springborg,
         J. Chem. Phys. {\bf 121}, 12179 (2004).             

\bibitem{Pines} D. Pines, Phys. Rev. {\bf 92}, 626 (1953).

\bibitem{GMB} M. Gell-Mann and K. A. Brueckner, 
                    Phys. Rev. {\bf 106}, 364 (1957).

\bibitem{Hedin} L. Hedin, Phys. Rev. {\bf 139}, A796 (1965).

\bibitem{PT2004} J. P. Perdew, J. Tao, V. N. Staroverov and G. E. Scuseria,
                   J. Chem. Phys. {\bf 120}, 6898 (2004).

\bibitem{WIv} L. C. Wilson and S. Ivanov, 
              Int. J. Quantum Chem. {\bf 69}, 523 (1998).

\bibitem{FUT} C. Filippi, C. J. Umrigar and M. Taut,
              J. Chem. Phys. {\bf 100}, 1290 (1994).
               

\end{thebibliography}
\end{document}